\documentstyle[sprocl]{article}

\def \MSbar {\vbox{\hrule\kern 1pt\hbox{\rm MS}}}
\def \sMSbar {\vbox{\hrule\kern 1pt\hbox{\smallrm MS}}}

\def\lesssim{\ \hbox{\raise 2pt \hbox{$<$} \kern -11pt
                     \lower 3pt \hbox{$\sim$}}\ }

\input epsf.tex
\def\DESepsf(#1 width #2){\epsfxsize=#2 \epsfbox{#1}}

\begin{document}

\title{QCD CALCULATIONS BY NUMERICAL INTEGRATION}
\author{ Davison E.\ Soper}
\address{
Institute of Theoretical Science, University of Oregon\\
Eugene, OR 97403, USA \\
E-mail: soper@bovine.uoregon.edu \\
{\it and}\\
Theoretical Physics Division, CERN\\
CH-1211 Geneva 23, Switzerland
}
\maketitle

\abstracts{
Calculations in Quantum Chromodynamics are typically performed using a
method pioneered by Ellis, Ross and Terrano in 1981. In this method, one
combines numerical integrations over the momenta of final state
particles with analytical integrations over the momenta of virtual
particles. I discuss a more flexible method in which one performs all of
the integrations numerically.
}

%
\section{Introduction}

This talk concerns a method for performing perturbative calculations in
Quantum Chromodynamics and other quantum field
theories.~\cite{beowulfprl}  The method may prove useful for cross
sections in which one measures something about the hadronic final state,
so that one cannot use the powerful but very special \hbox{methods} that
apply to completely inclusive quantities like the total cross section
for $e^+ e^- \to {\it  hadrons}$. Examples of such cross sections
include jet cross sections in hadron-hadron and lepton-hadron scattering
and in $e^+ e^- \to {\it  hadrons}$.

Many calculations of this kind have been carried out at next-to-leading
order in perturbation theory, resulting in an impressive confrontation
between theory and experiment. These calculations have been based on a
method introduced by Ellis, Ross, and Terrano~\cite{ERT} in the context
of $e^+ e^- \to {\it  hadrons}$. Stated in the simplest terms, one
performs some integrations over momenta $\vec p_i$ analytically, others
numerically. I shall argue that it is possible instead to do all of
these integrations numerically. This has some advantages, principally in
the flexibility that it allows.

In this talk, I address specifically the calculation of three-jet-like
infrared safe observables in $e^+ e^- \to {\it  hadrons}$ at
next-to-leading order, that is order $\alpha_s^2$. Examples of such
observables include the thrust distribution, the fraction of events that
have three jets, and the energy-energy correlation function. There are,
of course, several existing computer programs that can calculate any
such quantity, the earliest of these being that of Kunszt and
Nason,~\cite{KN} which was developed in 1989. What is new in this talk
is not the result achieved, but the method of calculation.

\section{The quantity to be calculated.}

The order $\alpha_s^2$ contribution to the observable being calculated
has the form
\begin{eqnarray}
\sigma^{[2]} &=&
{1 \over 3!}
\int d\vec k_1 d\vec k_2 d\vec k_3\
{d \sigma^{[2]}_3 \over d\vec k_1 d\vec k_2 d\vec k_3}\
{\cal S}_3(\vec k_1,\vec k_2,\vec k_3)
\label{start}\\
&&
+
{1 \over 4!}
\int d\vec k_1 d\vec k_2 d\vec k_3 d\vec k_4\
{d \sigma^{[2]}_4 \over d\vec k_1 d\vec k_2 d\vec k_3 d\vec k_4}\
{\cal S}_4(\vec k_1,\vec k_2,\vec k_3,\vec k_4).
\nonumber
\end{eqnarray}
There are, of course, infrared divergences associated with
Eq.~(\ref{start}); we understand that an infrared cutoff has been
supplied. In Eq.~(\ref{start}), the $d\sigma^{[2]}_n$ are the  order
$\alpha_s^2$ contributions to the parton level cross section, calculated
with zero quark masses. Each contains momentum and energy conserving
delta functions. The $d \sigma^{[2]}_n$ include ultraviolet
renormalization in the \MSbar\ scheme. The functions $\cal S$ describe
the measurable quantity to be calculated. Since we calculate a
``three-jet'' quantity, ${\cal S}_2 = 0$. The measurement, as
specified~\cite{KS} by the functions ${\cal S}_n$, is to be infrared
safe: the ${\cal S}_n$ are smooth functions of the parton momenta and
\begin{equation}
{\cal S}_{n+1}(\vec k_1,\dots,\lambda \vec k_n,(1-\lambda)\vec k_n)
= 
{\cal S}_{n}(\vec k_1,\dots, \vec k_n)
\end{equation}
for $0\le \lambda <1$. That is, collinear splittings and soft particles
do not affect the measurement.

It is convenient to calculate a quantity that is dimensionless. Let us,
therefore, define the functions ${\cal S}_n$ so that they are
dimensionless and eliminate the remaining dimensionality in the problem
by dividing by $\sigma_0$, the total $e^+ e^-$ cross section at the Born
level. We also remove the factor of $(\alpha_s / \pi)^2$. Thus, we
calculate
\begin{equation}
{\cal I} = {\sigma^{[2]} \over \sigma_0\ (\alpha_s/\pi)^2}.
\label{rtsintegral}
\end{equation}

\section{Setting up the calculation}

We note that ${\cal I}$ is a function of the c.m.\ energy $\sqrt s$ and
the $\overline{\rm MS}$ renormalization scale $\mu$. We will choose
$\mu$ to be proportional to $\sqrt s$: $\mu = A_{UV} \sqrt s$. Then
${\cal I}$ depends on $A$. But, because it is dimensionless, it is
independent of $\sqrt s$. This allows us to write
\begin{equation}
{\cal I} = \int_0^\infty d \sqrt s\ h(\sqrt s)\ 
{\cal I}(A_{UV},\sqrt s),
\end{equation}
where $h$ is any function with
\begin{equation}
\int_0^\infty d \sqrt s\ h(\sqrt s) = 1.
\end{equation}
The integration over $\sqrt s$ eliminates the energy conserving delta
function in $\cal I$. This is important in allowing  cancellations of
infrared singularities between contributions with 3- and 4-particle
intermediate states to take place. The physical meaning is that, by
smearing in the energy $\sqrt s$, we force the time variables in the two
current operators that create the hadronic state to be within $1/\sqrt
s$ of each other. Thus we have a truly short distance problem.

\section{The Ellis, Ross, and Terrano method}

\begin{figure}
\centerline{\DESepsf(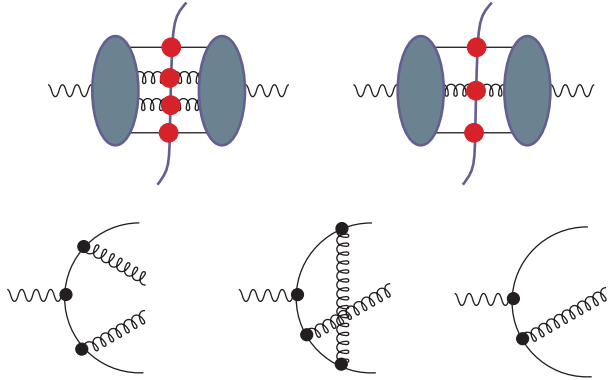 width 8 cm)}
\medskip

\caption{Illustration of the method of Ellis, Ross and Terrano. The dots
on the parton lines entering the final state represent the measurement
function. Some of the contributions to the amplitudes are illustrated in
the lower part of the figure. For four particle intermediate states, one
multiplies tree amplitudes. For three particle intermediate states, one
multiplies a tree amplitude by a loop amplitude. }
\label{ERT.fig}
\end{figure}

I now describe how one would calculate $\cal I$ using the
Ellis-Ross-Terrano method. Each partonic cross section in
Eq.~(\ref{start}) can be expressed as an amplitude times a complex
conjugate amplitude, as illustrated in Fig.~\ref{ERT.fig}. One must
calculate the amplitudes in $4 - 2 \epsilon$ dimensions. (In the case of
the process $e^+e^- \to {\it hadrons}$, this calculation was performed
by  Ellis, Ross, and Terrano.~\cite{ERT}) For tree diagrams, the
calculation is straightforward. For loop diagrams, this involves an
integration, which is performed analytically. The integrals are
divergent in four dimensions, so one obtains divergent terms
proportional to $1/\epsilon^2$ and $1/ \epsilon$ in addition to terms
that are finite as $\epsilon \to 0$. Having the amplitudes and complex
conjugate amplitudes, one must now multiply by the functions ${\cal
S}_n$ and integrate over the final state parton momenta. These
integrations are too complicated to perform analytically, so one must
use numerical methods. Unfortunately, the integrals are divergent at
$\epsilon = 0$. Thus one must split the integrals into two parts. One
part can be divergent at $\epsilon = 0$ but must be simple enough to
calculate analytically. The other part can be complicated, but must be
convergent at $\epsilon = 0$. One calculates the simple, divergent part
and cancels the $1/\epsilon^2$ and $1/ \epsilon$ pole terms against the
pole terms coming from the virtual loop diagrams. This leaves the
complicated, convergent integration to be performed numerically.

This method is a little bit cumbersome, but it works and has been
enormously successful. However it has proven to be difficult to apply
the method in the case of two virtual loops. Even with one virtual loop,
the method is not very flexible. In modern implementions of the Ellis-
Ross-Terrano method, one can easily choose the functions ${\cal S}_n$, but
any other modification of the integrand requires one to recalculate the
amplitudes, and the modification must be simple enough that one can
calculate the amplitudes in closed form. 

\section{Calculation by numerical integration}

\begin{figure}
\centerline{\DESepsf(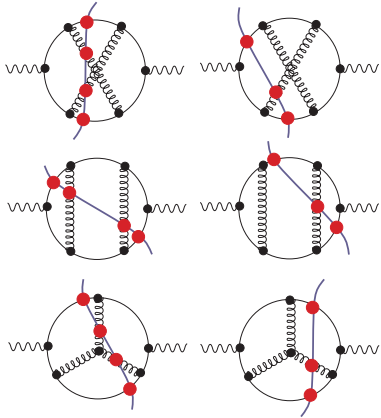 width 5 cm)}
\medskip
\caption{Some cut Feynman diagrams that contribute to $e^+e^-
\to {\it hadrons}$.}
\label{cutdiagrams}
\end{figure}

Let us, therefore, inquire whether there is any other way that one might
perform such a calculation. We note that the quantity $\cal I$ can be
expressed in terms of cut Feynman diagrams, as in
Fig.~\ref{cutdiagrams}. Each diagram is a three loop diagram, so we have
integrations over loop momenta $\ell_1^\mu$, $\ell_2^\mu$ and
$\ell_3^\mu$. 

\subsection{Performing the energy integrations}

We first perform the energy integrations. For the graphs in which four
parton lines cross the cut, there are four mass-shell delta functions
$\delta (k_J^2)$. These delta functions eliminate the three energy
integrals over $\ell_1^0$, $\ell_2^0$, and $\ell_3^0$ as well as the
integral (\ref{rtsintegral}) over $\sqrt s$. For the graphs in which
three parton lines cross the cut, we can eliminate the integration over
$\sqrt s$ and two of the $\ell_J^0$ integrals. One integral over the
energy $E$ in the virtual loop remains. The integrand contains a product
of factors
\begin{equation}
{ i \over E - Q_J^0 - \omega_J + i\epsilon}\,
{ 1 \over E - Q_J^0 + \omega_J - i\epsilon}\ \ ,
\end{equation}
where $E - Q^0_J$ is the energy carried by the $J$th propagator around
the loop and $\omega_J$ is the absolute value of the momentum carried on
that propagator. We perform the integration by closing the integration
contour in the lower half $E$ plane. This leads to $n$ terms for a
virtual $n$ point subgraph. In the $J$th term, the propagator energy $E
- Q^0_J$ is set equal to the corresponding $\omega_J$ and there is a
factor $1/(2\omega_J)$. Note that the entire process of performing the
energy integrals amounts to some simple algebraic substitutions.

\subsection{Cancellations of pinch singularities}

Let us denote the contribution to $\cal I$ from the cut $C$ of graph $G$
by ${\cal I}(G,C)$. This contribution has the form
\begin{equation}
{\cal I}(G,C) = \int d^3\vec \ell_1\,d^3\vec \ell_2\,d^3\vec \ell_3\
g(G,C;\ell),
\end{equation}
where $\ell$ denotes the set of loop momenta $\{\vec\ell_1, \vec\ell_2,
\vec\ell_3\}$. The functions $g$ have some singularities, called pinch
singularities, that cannot be avoided by deforming the integration
contour and some non-pinch singularities that can be avoided by a
contour deformation. 

Let us consider the pinch singularities, following the analysis of
Sterman.~\cite{sterman}  The pinch singularities occur when one parton
branches into two partons with collinear momenta or when one parton
momentum goes to zero. These singularities can lead to logarithmic
divergences in the corresponding integral. (There is a complication
associated with the $q^\mu q^\nu$ terms in gluon self-energy subgraphs
but I ignore this complication in this talk.) If we were to calculate
the total cross section by using measurement functions ${\cal S}_n = 1$
in $g$, then the singularities would cancel between the functions $g$
associated with the various cuts $C$ of the same graph $G$. The
underlying reason is unitarity. Now in our case of measured shape
variables, the values of ${\cal S}_n$ corresponding to different cuts
$C$ are different. This would ruin the cancellation, except that just at
the collinear or soft points the functions ${\cal S}_n$ match.  Thus the
singularities present in the individual $g(G,C;\ell)$ cancel in the sum,
$\sum_C g(G,C;\ell)$. To be precise, the cancellation reduces the
strength of the singularity from a strength sufficient to give a
logarithmically divergent integral to one that gives a convergent
integral.

\subsection{Avoiding the non-pinch singularities}

The function $g$ also has singularities that can be avoided by deforming
the integration contours, the {\it non-pinch} singularities. These
singularities do not cancel, but the Feynman rules provide an
$i\epsilon$ prescription that tells us that we should deform the $\vec
\ell$ integration contour into the complex $\vec \ell$ plane so as to
avoid the singularity. Here deforming the contour means replacing $\vec
\ell$ by a complex vector $\vec \ell + i \vec \kappa$. Then one simply
chooses the imaginary part, $\vec \kappa$, of the loop momentum as a
function of the real part, $\vec \ell$, and supplies the appropriate
jacobian $\cal J$. Then
\begin{equation}
{\cal I}(G,C) = \int d^3\vec \ell_1\,d^3\vec \ell_2\,d^3\vec \ell_3\
{\cal J}(G,C;\ell)\
g(G,C;\ell + i\kappa(G,C;\ell)).
\end{equation}

\subsection{The integration}

With the contours
appropriately chosen, the integral
\begin{equation}
{\cal I}(G) = \int d^3\vec \ell_1\,d^3\vec \ell_2\,d^3\vec \ell_3\
\sum_C\,
{\cal J}(G,C;\ell)\
g(G,C;\ell + i\kappa(G,C;\ell))
\label{master}
\end{equation}
is finite. One can simply compute it by Monte Carlo integration after
removing from the integration tiny regions near the collinear and soft
singular points, where roundoff errors spoil the cancellation of the
individual contributions.

Note the significance of putting the summation over cuts inside the
integral. When we sum over cuts for a given point in the space of loop
momenta, the soft and collinear divergences cancel because the
cancellation is built into the Feynman rules. If we were to put the sum
over cuts outside the integration, as in the Ellis-Ross-Terrano method,
then the individual integrals would be divergent. The calculation would
thereby be rendered more difficult.

\section{Additional points}

Eq.~(\ref{master}) represents the main point of this talk.  There are
two additional points of principle that I discuss below:
renormalization and the special treatment required for self-energy
subdiagrams.

\subsection{Renormalization}

\begin{figure}
\centerline{\DESepsf(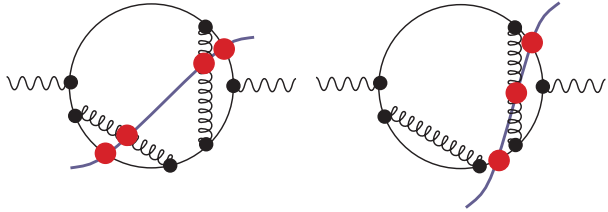 width 8 cm)}
\medskip
\caption{Two cuts of a diagram with a self-energy subgraph.}
\label{selfenergy.fig}
\end{figure}

Some of the virtual loop subgraphs are ultraviolet divergent. Normally,
one would renormalize these subgraphs by performing loop integrations in
$4-2\epsilon$ space-time dimensions and subtracting the resulting pole
term, $c/\epsilon$. Clearly that is not appropriate in a numerical
integration. However, one can subtract instead an integral in 4
space-time dimensions such that, in the region of large loop momenta,
the integrand of the subtraction term matches the integrand of the
subdiagram in question. The integrand of the subtraction term can depend
on a mass parameter $\mu$ in such a way that the subtraction term has no
infrared singularities. (For example, in the simplest case one can have
denominators of the form $1/[\vec\ell^2 + \mu^2]$.) Then, one can easily
arrange the definition so that this subtraction has exactly the same
effect as the conventional \MSbar\ subtraction with scale parameter
$\mu$.

\subsection{Self-energy subgraphs}

Virtual self-energy subgraphs, as in the right hand diagram in
Fig.~\ref{selfenergy.fig}, require a special treatment. Consider a quark
self-energy subdiagram $-i\Sigma$ with one adjoining virtual propagator
and one adjoining cut propagator. This combination really represents a
field strength renormalization for the quark field, and is interpreted
as
\begin{equation}
{ 1 \over 2}
\left[{ \rlap{/}q \over q^2}\,\Sigma(q)\,\rlap{/}q\right]_{q^2 = 0}\, 2\pi 
\delta(q^2).
\end{equation}
In order to take the $q^2 \to 0 $ limit here while at the same time
maintaining the cancellation of collinear divergences, we write
\begin{equation}
{ \rlap{/}q\,\Sigma(q)\,\rlap{/}q \over q^2}
= - { g^2 \over (2\pi)^3}\,C_F
\int d^3\vec\ell\
{ q^0 |\vec k_+|\gamma^0 - (|\vec k_+|+|\vec k_-|)\vec k_+\cdot \vec \gamma
\over |\vec k_+| |\vec k_-|\left[(|\vec k_+|+|\vec k_-|)^2 - (q^0)^2\right]},
\label{dispersion}
\end{equation}
where $\vec k_\pm ={ 1 \over 2}\vec q \pm \vec \ell$. This expression is
obtained by writing the left-hand side as a dispersive integral with the
cut self-energy graph appearing as the discontinuity. When the virtual
self-energy is written in this form, the $q^2 \to 0$ limit is smooth
and, in addition, the cancellation between the two cut diagrams in
Fig.~\ref{selfenergy.fig} in the collinear limit $\vec \ell \propto \vec
q$  is manifest. It should be noted that the integral in
Eq.~(\ref{dispersion}) is ultraviolet divergent and requires
renormalization, which can be performed with an {\it ad hoc} subtraction
as described above.

One may expect that the representation of virtual propagator corrections
in terms of the cut propagator will prove to be convenient in future
modifications of the method described here. One may want to make
modifications to the gluon propagator, in particular, in order to
implement a running strong coupling and to insert models for the long
distance propagation of gluons. In addition, one may want to modify
propagators so that partons can enter the final state with $q^2 >0$ so
as to mesh an order $\alpha_s^2$ perturbative calculation with a parton
shower Monte Carlo program.  In either case, the modified virtual gluon
propagator should be related by a dispersive representation to the
modified final state created by an outgoing gluon.

\section{An example at order $\alpha_s$}

I have argued that a completely numerical calculation of the $e^+e^-$
observables considered here is possible in principle. The question now
arises whether the principle can be put into operation in an actual
computer program.

\begin{figure}
\centerline{\DESepsf(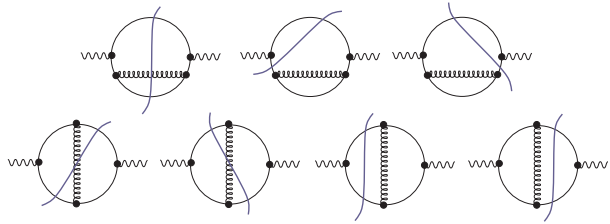 width 8 cm)}
\medskip
\caption{The cut diagrams for the total cross section for  $e^+e^- \to
hadrons$ at order $\alpha_s$}
\label{twoloop.fig}
\end{figure}

Consider first a simple example at one lower order in $\alpha_s$: the
total cross section for $e^+e^- \to hadrons$ at order $\alpha_s$. The
cut diagrams are shown in Fig.~\ref{twoloop.fig}.

Let us define $A_1(\mu)$ to be the sum of the top graphs in
Fig.~\ref{twoloop.fig} and $A_2(\mu)$ to be the sum of the bottom
graphs, so that
\begin{equation}
\sigma = \sigma_0 \left\{1 + \left[A_1(\mu) + A_2(\mu)\right]
 {\alpha_s \over \pi}\right\}.
\end{equation}
Then a simple analytical calculation shows that
\begin{equation}
A_1(\mu) = -{5 \over 3} - {4 \over 3}\ln(\mu/\sqrt s)\,,
\hskip 0.5 cm
A_2(\mu) =  {8 \over 3} + {4 \over 3}\ln(\mu/\sqrt s)\,.
\end{equation}

\begin{figure}
\centerline{\DESepsf(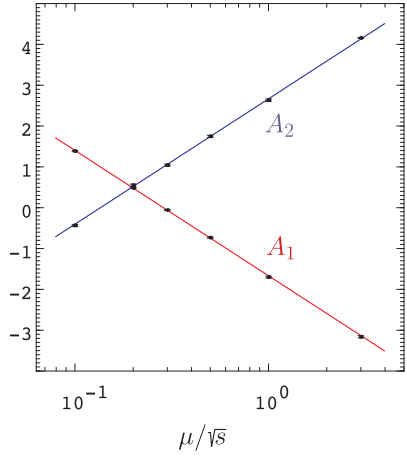 width 6 cm)}
\medskip
\caption{Result of the numerical calculation of the two pieces of the
total cross section for  $e^+e^- \to hadrons$ at order $\alpha_s$. The
numerically calculated functions $A_1(\mu)$ and $A_2(\mu)$ are plotted versus
the renormalization scale. The lines are the analytical results.}
\label{answer2.fig}
\end{figure}

I have constructed a demonstration computer program to calculate
$A_1(\mu)$ and $A_2(\mu)$ numerically along the lines described above.
This example has UV renormalization, propagators to be treated
dispersively, singularities to be avoided by deforming the contour, soft
singularities that must cancel, and collinear singularities that must
cancel. Thus it is not trivial from a numerical point of view. The
results are shown in Fig.~\ref{answer2.fig}. Evidently, the method
works.

\section{An example at order $\alpha_s^2$}
 
\begin{figure}
\centerline{\DESepsf(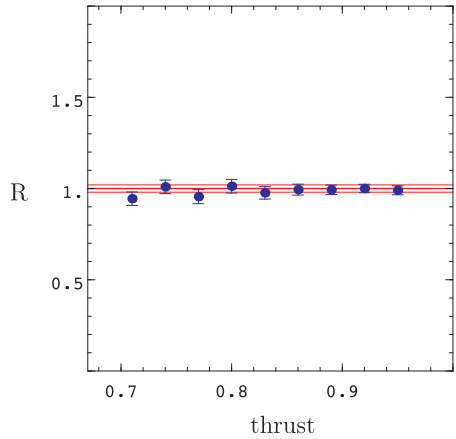 width 6 cm)}
\caption{Ratio $R$ of the order $\alpha_s^2$ contribution to the thrust
distribution as calculated here to the same quantity as calculated by
Kunszt and Nason.~\protect\cite{KN} The horizontal lines represent the
expected result, 1, with an error band based on the quoted errors in the
results of Kunszt and Nason and the uncertainties in fitting a smooth
function to these results. The error bars on the computed points
represent statistical errors. There is some correlation expected between
neighboring points.}
\label{answer}
\end{figure}

In order to further test the methods described here, I have constructed
a demonstration computer program~\cite{beowulf} to calculate
three-jet-like observables for $e^+e^-$ annihilation at order
$\alpha_s^2$. I have used the program to calculate $d\sigma^{[2]}/ d T$,
the order $\alpha_s^2$ contribution to the thrust distribution. More
precisely, I have calculated the ratio $R(T)$ of $d\sigma^{[2]}/ d T$ to
$[d\sigma^{[2]}/ d T]_{\rm KN}$, where where $[d\sigma^{[2]}/ d T]_{\rm
KN}$  is a fit to the tabulated results for $d\sigma^{[2]}/ d T$ as
given by Kunszt and Nason.~\cite{KN} In the range $0.71<T<0.95$, the
function $[d\sigma^{[2]}/ d T]_{\rm KN}$ varies by about a factor of 30.
The ratio $R(T)$ should be 1. The results are reported in
Fig.~\ref{answer}. Again, it appears that the numerical method of
calculation works in a practical sense.

\section{Outlook}
 
I am working to test the code of the demonstration program. Indeed,
between the time of the RADCOR98 conference and the time of this
writing, I have found and corrected a small error in the program logic
which resulted in an error in the calculation of about 1\%. With this
bug corrected, the results produced by the current version of the
program are better than those shown in  Fig.~\ref{answer}. After testing
the code, I will have to document the code and publish a detailed
description of the algorithms used.

With this code in hand, or with improved code from other authors, I
anticipate attacks on more difficult problems than the problem discussed
here. It remains to be seen for what problems the completely numerical
method will prove to be more powerful than the analytical/numerical
method that has served us so well up to now.

\section*{Acknowledgment}

This work was supported by the U.\ S.\ Department of Energy.


\section*{References}
\begin{thebibliography}{99}

\bibitem{beowulfprl} D.~E.~Soper, 
Phys\ .Rev.\ Lett.\ {\bf 81}, 2638 (1998).

\bibitem{ERT} R.~K.~Ellis, D.~A.~Ross, and A.~E.~Terrano, Nucl.\ Phys.\ {\bf
B178}, 421 (1981).

\bibitem{KN} Z.~Kunszt, P.~Nason, G.~ Marchesini and B.~R.~Webber in {\it Z
Physics at LEP1}, Vol.~1, edited by B.~Altarelli,
R.~Kleiss ad C.~Verzegnassi (CERN, Geneva, 1989), p.~373.

\bibitem{KS} Z.~Kunszt and D.~E.~Soper, Phys.\ Rev.\ D {\bf 46}, 192 (1992).

\bibitem{sterman} G.~Sterman, Phys.\ Rev.\ D {\bf 17}, 2773, 2789 (1978).


\bibitem{beowulf} The code used for Fig.~\ref{answer}, {\it beowulf} Version
0.7, is available at
http://zebu.uoregon.edu/\~{}soper/beowulf.html.

\end{thebibliography}
\end{document}